\documentclass[twocolumn,showpacs,preprintnumbers,amsmath,amssymb,prb,letterpaper]{revtex4}
\usepackage{graphicx}
\usepackage{comment}

\newcommand{\bmat}[4]{
\begin{bmatrix}
#1 & #2\\
#3 & #4\\
\end{bmatrix}
}
\newcommand{\paren}[1]{\left(#1\right)}

\newcommand{\coip}[3]{\left<#1\left|#2\right|#3\right>}

\newcommand{\tr}{\text{tr}\,}

\renewcommand{\exp}[1]{\left<#1\right>}
\newcommand{\mexp}[1]{\exp{\mathcal{#1}}}

\begin{document}

\bibliographystyle{apsrev}

\author{Gregory M. Crosswhite}
\email{gcross@phys.washington.edu}
\affiliation{Department of Physics; University of Washington; Seattle, WA 98185, U.S.A.}

\author{A.\ C.\ Doherty} 
\affiliation{School of Physical Sciences; University of Queensland; Brisbane, QLD 4072, Australia}

\author{Guifr\'e Vidal}
\affiliation{School of Physical Sciences; University of Queensland; Brisbane, QLD 4072, Australia}

\title{Applying matrix product operators to model systems with long-range interactions}


\begin{abstract}
An algorithm is presented which computes a translationally invariant matrix product state approximation of the ground state of an infinite 1D system;  it does this by embedding sites into an approximation of the infinite ``environment'' of the chain, allowing the sites to relax, and then merging them with the environment in order to refine the approximation.  By making use of matrix product operators, our approach is able to directly model any long-range interaction that can be systematically approximated by a series of decaying exponentials.  We apply these techniques to compute the ground state of the Haldane-Shastry model and present results.
\end{abstract}

\date{\today}

\maketitle

\tableofcontents

\section{Introduction}

Typical macroscopic quantities of material contain particules numbering on the order of $10^{26}$.  The only hope of modeling such systems is to come up with a representation that is relatively small but nonetheless able to capture essential properties.  For quantum systems this is particularly difficult because the state space grows exponentially with the size of the system, so there is a large amount of information which needs to be disposed.

One approach is to start with a small system and then build it up, using a renormalization transformation at each step to project out ``undesirable'' states and keep the representation small.  The key is to get one's criteria for ``undesirable'' correct.  When studying low-energy behavior of a system, it turns out that the best selection criteria is not the energy (unless one is solving the Kondo problem \cite{RevModPhys.47.773}), but the weight in a density matrix, which incorporates interactions with a surrounding environment.  This idea, first proposed by White \cite{White:1992ys} in his \emph{density matrix renormalization group} (DMRG) algorithm, has proven very successful in building effective but compact representations of large one-dimensional (1D) systems.  (See Ref. \onlinecite{Schollwock:2005ul} for an excellent review of this subject.)  
However, it has not worked as well as desired in modeling systems with long-range interactions, and it does not produce manifestly translationally invariant representations of translationally invariant states.  One can improve on both these fronts by working in momentum space rather than in real space \cite{PhysRevB.53.R10445,PhysRevB.65.165114}, but the computation is more costly as the transformation to momentum space introduces many additional operators that need to be renormalized at each step, and it can break symmetries that had been present in real space.  Also, none of these methods are particularly effective in two or more dimensions.

An alternative approach is inspired by observing that the DMRG
technique converges to a \emph{matrix product state} (MPS).
\cite{Ostlund:1995uq}  Rather than building up progressively larger
systems, the iTEBD algorithm\cite{cond-mat/0605597} assumes a priori
that the system takes an infinite translationally invariant
MPS form, and then uses imaginary time evolution
to converge to the ground state.  This approach has the advantage that it
obtains a very compact representation -- in particular, it dispenses
with the need to form renormalized versions of operators.
Furthermore, it admits a natural extension to two dimensions
\cite{cond-mat/0703788} by using \emph{projected entangled-pair
  states} (PEPS) \cite{cond-mat/0407066}, a two-dimensional
generalization of MPS.  However, it is only able to handle systems
with short-range interactions.

In this paper, we present an algorithm which also computes a translationally invariant MPS representation of a ground state, but without being limited to short-range interactions.  Instead of simulating an evolution in imaginary time as in the iTEBD algorithm, our approach follows the spirit of the variational technique described in Ref. \onlinecite{cond-mat/0404706}, but enhanced to apply to infinite systems;  additionally, we build in the ability to work with arbitrary \emph{matrix product operators} \cite{cond-mat/0701428,caching}, which makes this algorithm naturally suited to handle systems with long-range interactions.

The intuition behind our approach is as follows.  Suppose that one were given an infinitely large, translationally invariant 1D quantum chain held at zero temperature.  If one were to add an additional site to this chain and allow the chain to relax, then one would expect that all of the old sites would remain unchanged, and the new site would change to match the rest.  If one could emulate the environment experienced by a single site in this infinite chain, then by embedding a site into this environment and allowing the system to relax, one would obtain a site that ``looks like'' all of the sites in our infinite chain, giving us a compact representation for the chain.

We note that this algorithm bears some similarity to the product wave
function renormalization group (PWFRG) \cite{Nishino:1995kx} in that
both algorithms have the same goal---to compute representations of
ground states for infinite systems---and both use the same underlying
matrix product structure to represent states \cite{pwfrg-mps-2}.  The
difference is that while the PWFRG approach seeks the infinite limit
by starting with a small system and progressively enlarging it, we
start from the very beginning with an \emph{infinite} system,
represented in terms of effective environments which we progressively
refine.  Furthermore, we incorporate matrix product operators into our
approach which allow us to model systems with long-range interactions.

The remainder of this paper shall be organized as follows.  First, we shall review the matrix product formalism, and show how it allows us to construct a representation of a site embedded in and entangled with an environment.  Second, we shall outline the operation of an algorithm which uses an iteration procedure to approximate the effective environment of a site in an infinite chain;  we shall then explain how one can obtain the expected values of operators from the output of this algorithm.  Third, we shall explain how to obtain matrix product factorizations of Hamiltonians with exponentially decaying interactions.  Finally, we shall pull all of these ideas together and show how they can be applied to obtain an accurate MPS representation of the ground state of the Haldane-Shastry model---an exactly solvable model with long range interactions.

\section{Algorithms}

\label{algorithms}

\subsection{Finding a MPS representation of the ground state}

\label{main-algorithm}

We start by recalling the form of a matrix product state for a finite system \cite{Schollwock:2005ul}.  For a system with $n$ sites, a matrix product state with boundaries takes the form  \begin{eqnarray}
&&\mathcal{S}^{\alpha_1,\alpha_2,\dots,\alpha_n} = \nonumber \\
&&\quad\sum_{\{i_k\}} \paren{L^{\mathcal{S}}_0}_{i_1}\paren{S_1}^{\alpha_1}_{i_1 i_2} \paren{S_2}^{\alpha_2}_{i_2i_3}\dots \paren{S_n}^{\alpha_n}_{i_n i_{n+1}} \paren{R^{\mathcal{S}}_{n+1}}_{i_{n+1}}. \nonumber
\end{eqnarray}  The left-hand-side corresponds to the rank-n tensor describing our state;  each index $\alpha_k$ corresponds to a basis of a physical observable, such as the $z$-component of spin, at site $k$.  The rank-3 tensors $S_k$ are the site tensors, which have two kinds of indices:  the superscript index, which has dimension $d$ and corresponds to the actual physical observable, and the subscript indices, which have dimension $\chi$ and give information about entanglement between each site and its neighbors.  The vectors $L_0$ and $R_{n+1}$ give the boundary conditions.

The advantage of the matrix product representation is that it decomposes the quantum state into a collection of tensors, each of which is naturally associated with a site on the lattice.  Because of this, we shall see that we can ``zoom in'' on one site and then we only have to work with its corresponding tensor and a relatively small environment.  In this respect, it is almost as nice as a truly local description that would let us ignore all the other sites entirely--a fact which is remarkable given that it can still capture non-local properties resulting from entanglement.

Given this form, we now consider how to compute the expected values of operators.  As discussed in Ref. \onlinecite{cond-mat/0701428} and Ref. \onlinecite{caching}, operators of interest can typically be expressed as a matrix product operator, \begin{eqnarray}
&&\mathcal{O}^{(\alpha_1,\alpha_1'),(\alpha_2,\alpha_2'),\dots,(\alpha_n,\alpha_n')} = \nonumber \\
&&\quad\sum_{\{i_k\}} \paren{L_0^{\mathcal{O}}}_{i_1}\paren{O_1}^{\alpha_1,\alpha_1'}_{i_1 i_2}\dots \paren{O_n}^{\alpha_n,\alpha_n'}_{i_n i_{n+1}} \paren{R_{n+1}^{\mathcal{O}}}_{i_{n+1}} \label{matrix-product-operator}
\end{eqnarray}  Having decomposed both the state and the operator into a product of tensors associated with lattice sites, we likewise decompose the expected value into a product of tensors by combining the state and operator tensor at each site to define,  \begin{eqnarray}
\paren{E_k^{\exp{\mathcal{O}}}}_{i,j} \equiv \paren{E_k^{\exp{\mathcal{O}}}}_{(i',i'',i'''),(j',j'',j''')}:= \nonumber & \\
\sum_{\alpha_k,\alpha_k'} \paren{S_k^*}^{\alpha_k}_{i',j'}\paren{O_k}^{\alpha_k,\alpha_k'}_{i'',j''} \paren{S_k}^{\alpha_k'}_{i''',j'''}. \nonumber &
\end{eqnarray}  (Note that we have taken all of the left and right subscript indices and grouped them together.)  To complete our decomposition of the expected value, we shall also need to define the left and right boundaries, \begin{eqnarray}
&&\paren{L_0^{\exp{\mathcal{O}}}}_{i\equiv(i',i'',i''')}:= \paren{L_0^{\mathcal{S} *}}_{i'}\paren{L_0^{\mathcal{O}}}_{i''}\paren{L_0^{\mathcal{S}}}_{i'''}, \label{MPO-boundaries} \\
&&\paren{R_{n+1}^{\exp{\mathcal{O}}}}_{j\equiv(j',j'',j''')}:= \paren{R_{n+1}^{\mathcal{S}*}}_{j'}\paren{R_{n+1}^{\mathcal{O}}}_{j''}\paren{R_{n+1}^{\mathcal{S}}}_{j'''}. \nonumber
\end{eqnarray}  These boundary vectors can be thought of as defining an environment into which our system of $n$ sites
has been embedded.  With these tensors so defined, the expected value of the matrix product operator $\mathcal{O}$ with respect to the matrix product state $\mathcal{S}$ is given by $\coip{\mathcal{S}}{\mathcal{O}}{\mathcal{S}} = L_0^{\exp{\mathcal{O}}} \cdot E_1^{\exp{\mathcal{O}}} \cdot E_2^{\exp{\mathcal{O}}} \cdots E_n^{\exp{\mathcal{O}}} \cdot R_{n+1}^{\exp{\mathcal{O}}},$ where $\cdot$ indicates summation over the adjoining subscript indices.

Now we have almost arrived at the picture described in the introduction, except that we have an environment and multiple sites rather than a single site.  To compute the effective environment of a particular site in this chain, we absorb all of the sites surrounding it into the system environment by contracting the $E_k^{\mexp{O}}$ matrices into the left and right boundaries -- that is, we make the inductive definitions $L_k^{\exp{\mathcal{O}}}:= L_{k-1}^{\exp{\mathcal{O}}}\cdot E_k^{\exp{\mathcal{O}}}$ and $R_k:=E_k^{\exp{\mathcal{O}}}\cdot R_{k+1}^{\exp{\mathcal{O}}}$.  By doing this, we can write the expected value of the operator as a function of this site tensor,  \begin{eqnarray}
\label{fn-of-site-tensor}
&&\coip{\mathcal{S}}{\mathcal{O}}{\mathcal{S}}(S_k) \equiv S_k^* \circ M^{\exp{\mathcal{O}}_k} \circ S_k :=  \\
&&\quad= \sum_{\genfrac{}{}{0pt}{}{\alpha,\alpha',i',i''',}{j',j'''}}\paren{S_k^*}^\alpha_{i',j'} \paren{M^{\exp{\mathcal{O}}_k}}^{(\alpha),(\alpha')}_{(i',j'),(i''',j''')}\paren{S_k}^{\alpha'}_{i''',j'''} \nonumber
\end{eqnarray}  where $\circ$ denotes summation over the appropriate adjacent subscript and superscript indices, and  \begin{eqnarray}
\label{m-matrix-definition}
&&\paren{M^{\exp{\mathcal{O}}_k}}^{(\alpha),(\alpha')}_{(i',j'),(i''',j''')} := \nonumber\\
&&\quad\sum_{i'',j''}\paren{L_{k-1}^{\exp{\mathcal{O}}}}_{(i',i'',i''')} \paren{O_k}^{\alpha_k,\alpha_k'}_{i'',j''} \paren{R_{k+1}^{\exp{\mathcal{O}}}}_{(j',j'',j''')} \nonumber
\end{eqnarray}

We have now obtained an explicit means to compute the expected value
of an observable $\mathcal{O}$ as a function of the site tensor at position $k$
knowing only the environment of the site as given by
$L^{\mexp{O}}_{k-1}$ and $R^{\mexp{O}}_{k+1}$.  However, recall that
we want to be more than passive observers---we want to actively move
the system as close as possible to its ground state.  Thus, we now
want to vary the site tensor $S_k$ in order to minimize the energy of
the system.  If we let the Hamiltonian (matrix product) operator be
denoted by $\mathcal{H}$ and the identity operator by $\mathcal{I}$, then employing equation \eqref{fn-of-site-tensor}
we see that what we seek is the site tensor which minimizes the function
$$\mathcal{E}^k(S_k) := \frac{S_k^*
  \circ M^{\exp{\mathcal{H}}_k} \circ S_k}{S_k^* \circ
  M^{\exp{\mathcal{I}}_k} \circ S_k},$$
which gives us the (normalized) energy of the system as a function of only the site tensor at position $k$ (i.e., assuming that the environment has been frozen in place).  Since this is a Rayleigh quotient, computing the minimizer is equivalent to solving a generalized eigenvalue problem.

There is a subtlety in this procedure, however, which is that one needs to take steps to make sure that the normalization matrix $M^{\mexp{I}_k}$ is well-conditioned.  This can be done by imposing the ``right-normalization'' condition $\sum_{\alpha,i}\paren{S_{l}^*}^{\alpha}_{ij}\paren{S_{l}}^{\alpha}_{ij'}=\delta_{jj'}$ on all the sites to the left of $k$, and the ``left-normalization'' condition $\sum_{\alpha,j}\paren{S_{l}^*}^{\alpha}_{ij}\paren{S_{l}}^{\alpha}_{i'j}=\delta_{ii'}$ on all the sites to the right of $k$.  This ensures that the subscript indices connected to $S_k$ are orthonormal, which makes $M^{\mexp{I}_k}$ well-conditioned.  We shall discuss how this is done in our algorithm shortly;  for more information on how the normalization condition is used for finite-length systems, see Ref. \onlinecite{cond-mat/0404706}.

Now we have all of the ingredients that we need to build our algorithm:  a way of expressing a chain as a site tensor embedded in an environment, and a way to relax a system in this representation constrained so that only the site tensor (and not its environment) is changed.  However, up to this point we have been working with a finite system; in order to apply these ideas to infinite systems, we shall modify our notation slightly to replace position labels with \emph{iteration labels}.  That is, we shall let $L^{\mexp{O}}_k$ and $R^{\mexp{O}}_k$ denote the infinite environment
at iteration step $k$, and $S_k$ denote the inserted site at this iteration.  With this notation, the algorithm to find the ground state of a system is given in Table \ref{algorithm}. It is dominated by the costs of absorbing the site and operator tensors into the environment (in step 3b.iv, and in the Lanczos iteration in step 3a), which are respectively $O(cd\chi^3)$ and $O(c^2d^2\chi^2)$, with $c$ referring to the auxiliary dimension of the operator tensor.

\begin{table}
\begin{ruledtabular}
\begin{tabular}{p{0.075in}p{0.15in}p{0.1in}p{2.8in}}
1. & \multicolumn{3}{p{3.3in}}{Set $\chi=1$, and $L^{\mathcal{S}}=R^{\mathcal{S}}=1$.} \\
2. & \multicolumn{3}{p{3.3in}}{Compute $L_1^{\mexp{O}}$ and $R_1^{\mexp{O}}$ for the Hamiltonian
(matrix product)
operator $\mathcal{H}$ and the identity operator $\mathcal{I}$.
(The latter, of course, has trivial boundaries.)
} \\
3. & \multicolumn{3}{p{3.3in}}{Until convergence has been reached:} \\
   & (a) & \multicolumn{2}{p{2.9in}}{Use an eigenvalue solver (such as ARPACK \cite{arpack}) to obtain the minimal eigenvalue and corresponding eigenvector of the generalized eigenvalue problem $M^{\mexp{H}_k}S_k = \lambda M^{\mexp{I}_k} S_k.$  To accelerate convergence, feed in $S_{k-1}$ as a starting estimate for the eigenvector.
} \\
   & (b) & \multicolumn{2}{p{2.9in}}{If this is an odd-numbered step, then right-normalize $S_k$ and contract into the left boundary.  Specifically:} \\
   &     & i.  & Merge the superscript and first subscript index of $S_k$ to form a matrix, and compute the singular value decomposition (SVD), $U\cdot \Sigma\cdot V^\dagger$. \\
   &     & ii. & Set $\tilde S_k:=U\cdot V^\dagger$, and ungroup indices to return $\tilde S_k$ to its original rank-3 shape. \\
   &     & iii.& Compute $E_k^{\mexp{O}}$ for $\mathcal{H}$ and $\mathcal{I}$ using the normalized $\tilde S_k$. \\
   &     & iv. & Contract the site into the left boundary by setting $L_{k+1}^{\mexp{O}}:=L_{k}^{\mexp{O}}\cdot E^{\mexp{O}}_k$ and $R^{\mexp{O}}_{k+1}:=R_k^{\mexp{O}}$.  (This step ``absorbs'' the site into the environment.) \\
   & (c) & \multicolumn{2}{p{2.9in}}{If this is an even-numbered step, perform an analogous process, but \emph{left}-normalize $S_k$ by merging the superscript and \emph{second} index together before the SVD, and then contract the normalized site into the \emph{right}-boundary.} \\
4. & \multicolumn{3}{p{3.3in}}{If a better approximation to the ground state is desired, then increase $\chi$ for the system and repeat step 3.  Increasing the dimension of a subscript can be done without altering the state by multiplying the adjoining tensors by a $\chi\times (\chi+\Delta\chi)$ matrix and its inverse;  this allows one to build on the work of previous iterations (and hence accelerate convergence), rather than having to start from scratch with the new $\chi$.
}
\end{tabular}
\end{ruledtabular}
\caption{Algorithm to compute a (normalized) translationally invariant matrix product state representation of the ground state of an infinite chain.}
\label{algorithm}
\end{table}

We note here that this algorithm is unstable when applied directly to systems with anti-ferromagnetic interactions because ground states of such systems are not invariant under translations of one site.  Happily, since such systems are invariant under translations of \emph{two} sites, there is a simple fix:  work with blocks of two spins rather than one by setting $d=4$ and multiplying two of the operator tensors together to form a two-site operator tensor; \cite{tiabrefhack}  this only affects the inputs to the algorithm and does not require changing the algorithm itself.

\subsection{Computing expected values of operators}

\label{exp-algorithm}

The output of the algorithm of Sec. \ref{main-algorithm} is a normalized site tensor $\tilde S$ which gives a translationally invariant representation for the ground state of the system.  In order to make this useful, we need to have a way to obtain the expected value of operators from it.  Of course, for extensive observables the expected value will be infinite since we have an infinitely large system, so instead we seek the more useful quantity of the expected value per site.  We shall consider two cases of operators:  local operators and general matrix product operators.  The basic trick in both cases is to note that $\lim_{N\to\infty}\paren{E^{\mexp{O}}}^N \cdot \vec{v} = \Lambda^N \cdot \vec{v}$, where $\Lambda$ is a matrix in the maximal eigenspace of $E^{\mexp{O}}$ -- that is, in the infinite limit the action of the operator will converge to its action on the maximal eigenspace, since its action on all other eigenspaces will be negligible by comparison.  (Assuming, of course, that  $\Lambda \cdot \vec{v} \ne 0$.)

First we consider local operators, which can be expressed as sum of terms of the form $I^{\otimes\infty} \otimes O_1 \otimes O_2 \otimes \dots \otimes O_N \otimes I^{\otimes\infty}$, such as a magnetic field operator, $I^{\otimes\infty} \otimes Z\otimes I^{\otimes\infty}$, or a two-point correlator, $I^{\otimes\infty}\otimes Z\otimes I^{\otimes r}\otimes Z \otimes I^{\otimes\infty}$.  Since our system is translationally invariant, to compute the expected value per site of this operator we need only evaluate the expected value of one term in the sum.  Assuming that $E^{\mexp{I}}$ has a non-degenerate maximal eigenvalue, we have that for (almost) any vector $\vec{v}$, $\paren{E^{\mexp{I}}}^\infty\cdot \vec{v} \sim \vec{v}^R$ and $\vec{v}\cdot \paren{E^{\mexp{I}}}^\infty \sim \vec{v}^L$, where $\vec{v}^L$ and $\vec{v}^R$ are the respective left and right eigenvectors corresponding to the maximal eigenvalue.  Ergo, the expected value of one term of this operator (and thus the expected value per site) is given by  $$\frac{\vec{v}^L\cdot E^{O_1}\cdot E^{O_2}\dots E^{O_N}\cdot \vec{v}^R}{\vec{v}^L\cdot \paren{E^{\mexp{I}}}^N \cdot \vec{v}^R}.$$

Next we consider general matrix product operators.  We start by writing an expression for the expected value per site for a finite chain of length $N$, whose site tensors are copies of the (normalized) site tensor obtained from the variational algorithm in Sec. \ref{main-algorithm}.  Since the infinite chain has no boundaries, to compute the expected value of the finite chain we need to explicitly supply left and right boundaries, respectively $L^{\mathcal{S}}$ and $R^{\mathcal{S}}$, and so the expected value per site of $\mathcal{O}$ is a function of the boundaries given by $$\mathcal{E}^N(L^{\mathcal{S}},R^{\mathcal{S}}):=\frac{1}{N}\frac{L^{\mexp{O}} \cdot \paren{E^{\mexp{O}}}^N \cdot R^{\mexp{O}}}{L^{\mexp{I}} \cdot \paren{E^{\mexp{I}}}^N \cdot R^{\mexp{I}}},$$  where $L^{\mexp{O}}$, $R^{\mexp{O}}$, $L^{\mexp{I}}$, and $R^{\mexp{I}}$ are implicitly functions of $L^{\mathcal{S}}$ and $R^{\mathcal{S}}$ given by equation \eqref{MPO-boundaries}.  Assuming $\mathcal{O}$ corresponds to an extensive observable, we expect that $\lim_{N\to\infty} \mathcal{E}^N(L_{\mathcal{S}},R_{\mathcal{S}}) = \mexp{O}$ -- that is, in the infinite limit the expected value per site converges to some number $\mexp{O}$ which is independent of the boundaries.  With this physically reasonably assumption, we can reason about the structure of $E^{\mexp{O}}$ and $E^{\mexp{I}}$ to obtain an algorithm for computing $\mexp{O}$.

We first observe that since the maximal eigenvalue of $E^{\mexp{I}}$ is 1 (due to the normalization of the site tensor, $\tilde S$), so must be the maximal eigenvalue of $E^{\mexp{O}}$, since otherwise $\mathcal{E}^N$ would be exponential in $N$.  Furthermore, since $\mathcal{E}^N$ is linear in $N$ for large $N$, the maximal eigenspace of $E^{\mexp{I}}$ must have a Jordan block structure--that is, there must be a matrix $U\equiv[\,\vec{u}_1\,\,\vec{u}_2\,]$  with orthonormal columns $\vec{u}_i$ that provide a basis for this eigenspace such that $A:=U \cdot E^{\mexp{O}} \cdot U^\dagger = \bmat{1}{\alpha}{0}{1}.$ The matrix element $\alpha$ can be thought of as giving us the unnormalized expected value of $\mathcal{O}$ per site.  In order to normalize it, we observe that as $N\to\infty$,
$$
\begin{aligned}
\mathcal{E}^N&\to
\frac{1}{N}\frac{L^{\mexp{O}} \cdot \paren{\vec{u}_1 {\vec{u}_1}^\dagger + \vec{u}_2 {\vec{u}_2}^\dagger + N\alpha \vec{u}_1{\vec{u}_2}^\dagger}\cdot R^{\mexp{O}}}{L^{\mexp{I}} \cdot \paren{E^{\mexp{I}}}^N \cdot R^{\mexp{I}}} \\
& \to
\alpha\frac{L^{\mexp{O}} \cdot (\vec{u}_1 {\vec{u}_2}^\dagger) \cdot R^{\mexp{O}}}{L^{\mexp{I}} \cdot \paren{E^{\mexp{I}}}^N \cdot R^{\mexp{I}}}.
\end{aligned}$$
Since we expect $\mathcal{E}^N$ to be independent of the boundaries for large $N$, it must be the case that $$
\begin{aligned}
L^{\mexp{I}} \cdot \paren{E^{\mexp{I}}}^N \cdot R^{\mexp{I}}
&\to \beta \left[ L^{\mexp{O}} \cdot (\vec{u}_1{\vec{u}_2}^\dagger) \cdot R^{\mexp{O}}\right]\\
&\equiv \beta \left[L^{\mexp{I}} \cdot (\vec{v}^L_1 \vec{v}^{R\dagger}_2) \cdot R^{\mexp{I}}\right], \\
\end{aligned}
$$ where $\vec{v}^L_{k} :=  \vec{u}_k \cdot L^{\mathcal{O}} \equiv \sum_{i''} u_{k,(i',i'',i''')} L^{\mathcal{O}}_{i''}$, and $\vec{v}^R_{k} :=  \vec{u}_k \cdot R^{\mathcal{O}} \equiv \sum_{i''} u_{k,(i',i'',i''')} R^{\mathcal{O}}_{i''}$ -- that is, $\vec{v}^L_k$ and $\vec{v}^R_k$ are the projections of the eigenspace of $E^{\mexp{O}}$ into a space applicable to $E^{\mexp{I}}$ by dotting out the respective left and right boundaries of the matrix product operator $\mathcal{O}$; put another way, if we let $V^L:=[\,\vec{v}^L_1\,\,\vec{v}^L_2\,]$ and $V^R:=[\,\vec{v}^R_1\,\,\vec{v}^R_2\,]$, then the limiting action of $E^{\mexp{I}}$ in the (projected) eigenspace is given by $B:=V^L \cdot E^{\mexp{I}} \cdot {V^R}^\dagger = \bmat{0}{\beta}{0}{0}$.  The matrix element $\beta$ gives us the normalization constant, so that $\mexp{O}:=\lim_{N\to\infty}\mathcal{E}^N=\alpha/\beta$.

It remains to extract $\alpha$ and $\beta$ from the respective matrices $A$ and $B$.  Our eigenvalue solver is not guaranteed to give us a basis of the eigenspace that makes $A$ and $B$ have the nice form above, but happily it is easy to see that $\alpha = \sqrt{\tr(A\cdot A^\dagger)-2}$ and $\beta =\sqrt{\tr(B\cdot B^\dagger)}$, and these formulas are invariant under similarity transforms of $A$ and $B$ so they work independent of the basis we are given.  The algorithm we have derived in the preceding paragraphs is summarized in Table \ref{MPO-expectation-algorithm}.

\begin{table}
\begin{ruledtabular}
\begin{tabular}{p{0.075in}p{0.15in}p{0.1in}p{2.8in}}
1. & \multicolumn{3}{p{3.3in}}{Compute the maximal two-dimensional eigenspace of $E^{\mexp{O}}$---for example, by using ARPACK \cite{arpack} and requesting a Schur basis rather than a (non-existant) eigenvector basis---and store the basis vectors in the columns of a matrix $U$.} \\
2. & \multicolumn{3}{p{3.3in}}{Express the action of $E^{\mexp{O}}$ in this space by computing the $2\times 2$ matrix $A:=U\cdot E^{\mexp{O}}\cdot U^\dagger.$} \\
3. & \multicolumn{3}{p{3.3in}}{Extract the unnormalized expected value from this matrix by computing $\alpha:=\sqrt{\tr(A^\dagger\cdot A)-2}.$} \\
4. & \multicolumn{3}{p{3.3in}}{Project out the operator-dependent boundary conditions from the vectors in this eigenspace to obtain $V^L:=L^{\mathcal{O}}\cdot U\equiv \sum L^{\mathcal{O}}_{i''} U_{(i',i'',i'''),k}$ and $V^R:=R^{\mathcal{O}}\cdot U\equiv \sum R^{\mathcal{O}}_{i''} U_{(i',i'',i'''),k}$.} \\
5. & \multicolumn{3}{p{3.3in}}{Express the action of $E^{\mexp{I}}$ in this projected space by computing the $2\times 2$ matrix $B:={V^L} \cdot E^{\mexp{I}} \cdot {V^R}^\dagger$.} \\
6. & \multicolumn{3}{p{3.3in}}{Extract the normalization constant by computing $\beta:=\sqrt{\tr(B^\dagger\cdot B)}.$} \\
7. & \multicolumn{3}{p{3.3in}}{Obtain the (normalized) expected value by computing $\mexp{O}:=\alpha/\beta$.}
\end{tabular}
\end{ruledtabular}
\caption{Algorithm to compute $\mexp{O}$ for general matrix product operator $\mathcal{O}$.}
\label{MPO-expectation-algorithm}
\end{table}

\section{Matrix product factorization of exponentially decaying interactions}

\label{automaton}

In the algorithms of Sec. \ref{algorithms}, we assume that there is a matrix product representation of our Hamiltonian.  It has previously been shown that there are matrix product factorizations of Hamiltonians with short-range interactions.\cite{cond-mat/0701428,caching}  In this section, we extend these results to show that there are also matrix product factorizations of Hamiltonians with long-range exponentially decaying interactions as well.

So consider a Hamiltonian of an $N$-site system experiencing a sum of long-range exponentially decaying interactions,
$$\mathcal{H}:=\sum_{i+1+r+1+j=N} \paren{\sum_n \alpha_n \beta_n^r} \textbf{I}^{\otimes i}\otimes \textbf{X} \otimes \textbf{I}^{\otimes r}\otimes \textbf{X} \otimes \textbf{I}^{\otimes j},$$ where $\textbf{I}$ is the identity operator and $\textbf{X}$ is the Pauli operator $\sigma^X$.  We shall now employ the formalism in Ref. \onlinecite{caching} to obtain a factorization of this Hamiltonian.  We start by changing our interpretation of the Hamiltonian:  rather than thinking of it as a sum of tensor product terms with complex coefficients, we shall instead think of it as a function which maps arbitrary strings of \texttt{X} and \texttt{I} \emph{symbols} to complex numbers.  In our sum each term takes the form $\paren{\sum_n \alpha_n \beta_n^r} \textbf{I}^{\otimes i}\otimes \textbf{X} \otimes \textbf{I}^{\otimes r}\otimes \textbf{X} \otimes \textbf{I}^{\otimes j}$ with $i,j,r \ge 0$, so our corresponding function maps strings of the form $\texttt{I}^i\, \texttt{X}\, \texttt{I}^r\, \texttt{X}\, \texttt{I}^j$ to $\sum_n \alpha_n \beta_n^r$, and all other strings to zero.

We shall now construct a \emph{finite state automaton} which computes this function.  A finite state automaton can be thought of as a machine which reads through an input string and changes its state in response to each input symbol according to a set of transition rules.  The transitions are non-deterministic and weighted -- that is, the automaton can take many transitions simultaneously, and at the end of the string it outputs a number corresponding to the sum of the product of the weights along each sequence of transitions that it took.  The automaton starts on a designated \emph{initial} state;  if it does not end on a designated \emph{accept} state or it encounters a symbol without an associated transition, then it outputs a zero (overriding other weights).

\begin{figure}
\begin{center}
\framebox{\includegraphics[width=2.5in]{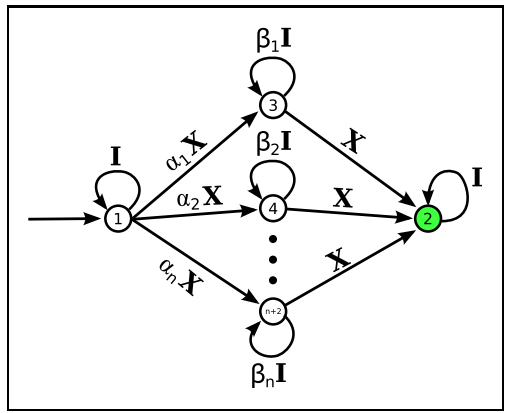}}
\caption{(color online) Finite state automata representation of a sum of exponentially decaying $XX$ interactions.}
\label{finite-state-automaton}
\end{center}
\end{figure}

The automaton which computes the function representing our operator is illustrated in Fig. \ref{finite-state-automaton}.  States are indicated by circles, and transition rules are indicated by arrows labeled with a symbol and a weight  (one if not otherwise specified).  An unconnected arrow designates the initial state (1), and shading designates the accept state (2).  A good way to think about what is going on is that terms are generated by each possible walk from state $1$ to state $2$.  So for example, by taking the path $1\!\!\!\to\!\!1\!\!\to\!\!3\!\!\to\!\!3\!\!\to\!\!3\!\!\to\!\!2$ we generate the term $(\alpha_1 \beta_1^2) \textbf{I} \otimes \textbf{X} \otimes \textbf{I}\otimes \textbf{I} \otimes \textbf{X}$;  summing over all walks for strings of the form $\texttt{I}\,\texttt{X}\,\texttt{I}^2\texttt{X}$ we obtain the desired coefficient $\sum_n \alpha_n \beta_n^2$.

From this automaton, we immediately obtain a matrix product operator representation.  The initial and accept states give us the values for the left and right boundaries:  $\paren{L^{\mathcal{O}}}_k=\delta_{1,k}$ and $\paren{R^{\mathcal{O}}}_k=\delta_{2,k}$.  The elements of the operator tensor, $O^{\textbf{A}}_{i,j}$, are given by the weight on the $i\to j$ transition with the symbol corresponding to operator $\textbf{A}$ (zero if no such transition exists);  for example, we have that $O^{\textbf{X}}_{1,3}=\alpha_1$.  To get a feeling for why this works, observe that a run of the automaton is equivalent to starting with a vector giving initial weights on the states, multiplying this vector some number of times by a transition matrix, and then dotting the result with a vector that filters out all but the weights on the accept states;  this procedure is exactly equivalent to the form of equation \eqref{matrix-product-operator}.

This process is easily extended to include terms with arbitrary spin coupling interactions, such as $\sigma^X\sigma^Y$, $\sigma^Y\sigma^Y$, $\sigma^Z\sigma^Z$, etc.  Furthermore, one can combine a sum of several such interactions into a single automaton by having them all share the same starting and ending states. 

\section{Results:  Haldane-Shastry Model}

Now we pull all of the ideas from the previous sections together and apply them to tackle the Haldane-Shatry model.\cite{PhysRevLett.60.635,PhysRevLett.60.639}  In the infinite limit this model is given by the Hamiltonian $\mathcal{H} = \sum_i \sum_r \vec{\sigma}_i\cdot\vec{\sigma}_{i+r}/r^2$,  which features an anti-ferromagnetic dipole interaction which falls off with the square of the distance between sites.  (Note that since this model features anti-ferromagnetic interactions, we need to work with blocks of two sites, as discussed at the end of section \ref{main-algorithm}.)  Although $\mathcal{H}$ cannot be expressed exactly as a matrix product operator, we can approximate it arbitrarily well by a sum of exponentially decaying interactions, $\mathcal{H}\approx  \sum_i \sum_r \vec{\sigma}_i\cdot\vec{\sigma}_{i+r}\paren{\sum_n \alpha_n \beta_n^{r-1}}$ (with $\alpha_n\in\mathbb{R}$ and $|\beta_n| \le 1$), which can be factored exactly using the technique in section \ref{automaton}.  Since there are three spin-coupling interactions, $\sigma^X_i\sigma^X_{i+r}$, $\sigma^Y_i\sigma^Y_{i+r}$, and $\sigma^Z_i\sigma^Z_{i+r}$, which we combine into a single automaton as discussed at the end of section \ref{automaton}, we obtain an automaton with a number of states equal to three times the number of terms in the expansion, $N$, plus two more for the starting and ending states;  this quantity gives us the size of the auxiliary dimension for the corresponding matrix product operator, $c=3N+2$.

It remains to find the coefficients in this expansion.  One approach is to numerically solve for the coefficients which minimize the sum of the squares of the difference between the approximation and the exact potential for distances up to some cut-off -- that is, to find the minimizer of the function,
$$f(\alpha_1,\beta_1,\dots,\alpha_N,\beta_N) = \sum_{i=1}^N \sum_{r=1}^{r_{\text{cutoff}}} \paren{\alpha_i \beta_i^{r-1} - \frac{1}{r^2}}^2,$$
where $r_{\text{cutoff}}$ should be chosen to be just beyond the maximum effective range of the approximation, since larger values of $r_{\text{cutoff}}$ result in a longer running time for the minimization without resulting in a better fit.

\begin{figure}
\includegraphics[width=\columnwidth]{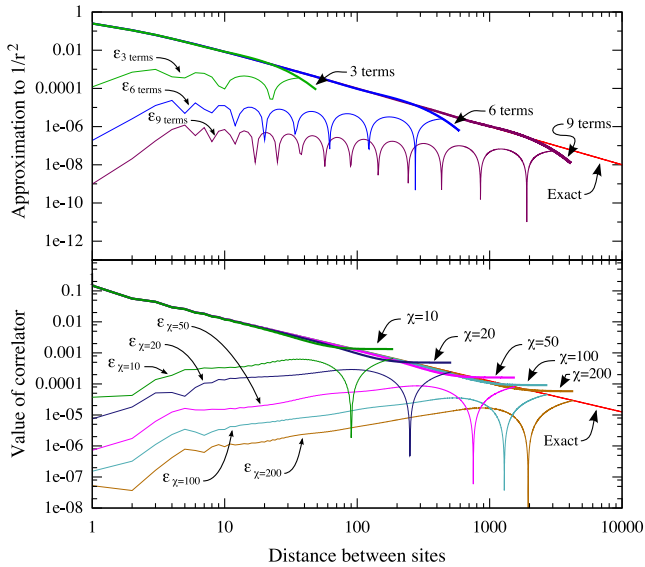}
\caption{(color online) (top) Decaying 3-, 6-, and 9-term exponential approximations to $1/r^2$ potential.  (bottom) Two-point correlator for selected values of $\chi$ using the the 9-term exponential approximation;  a residual of the correlator for each value of $\chi$ is plotted simultaneously and labeled by $\varepsilon_\chi$.}
\label{correlator}
\label{expansion}
\end{figure}

For our application of the algorithm, we used a nonlinear least-squares minimization routine from \texttt{MINPACK} to find coefficients for expansions with three, six, and nine terms;  the resulting approximate potentials are plotted along side the exact potential in Fig. \ref{expansion}.  The upper cutoff on $r$ was set to 10000 because, as can be seen in Fig. \ref{expansion}, this was just beyond the effective maximum range obtainable from a 9-term approximation.

\begin{figure}
\includegraphics[width=\columnwidth]{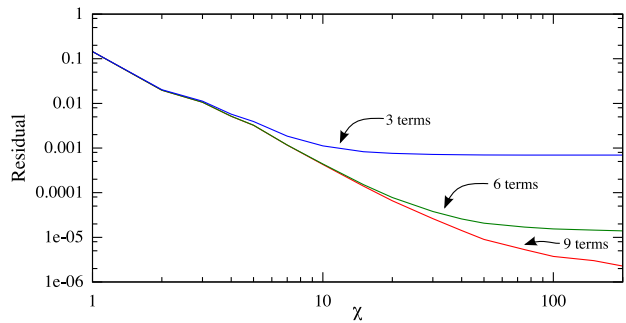}
\caption{(color online) Difference between the energy of the computed state and the energy of the exact ground state, plotted for each of the three exponential expansions that were employed.}
\label{residuals}
\end{figure}

Given this approximate matrix product factorization, we applied the algorithm in Table \ref{algorithm} to compute a translationally invariant matrix product state representation of the ground state for selected values of $\chi$, employing each of the three-, six-, and nine-term expansions.  The energy per site was computed using the algorithm in Table \ref{MPO-expectation-algorithm}, and compared to the exact value obtained from Ref. \onlinecite{PhysRevLett.60.639}.  The difference between these values (i.e., the residual) is plotted for each expansion as a function of $\chi$ in Fig. \ref{residuals}.  Note that the residuals for all three expansions agree up to some point, and then diverge to different ``floors''.  This is because at first the small value of $\chi$ is the dominating factor which limits the fidelity of the ground state, and then later as $\chi$ becomes large the finite number of terms in the exponential approximation becomes the dominating factor.

For the nine-term expansion, we also computed the two-point correlator -- that is, $C(r):=\exp{\sigma^X \otimes I^{\otimes (r-1)}\otimes\sigma^X}$ -- using the algorithm given in section \ref{exp-algorithm} for computing the expected value of a local operator.  The result for several values of $\chi$ is plotted in Fig. \ref{correlator} against the exact value from Ref. \onlinecite{PhysRevLett.60.639}.  Note that our approximation gets good agreement up to some length, after which it becomes a constant.  This is because the algorithm is attempting to approximate this correlator using a sum of decaying exponentials plus a constant term, analagous to how we used a sum of decaying exponentials to approximate the $1/r^2$ interactions;  by increasing $\chi$, we are increasing the number of terms available to track the correlator, which results in systematic improvement.

\section{Conclusions}

To summarize, in this paper we have presented an algorithm for
computing the ground state of infinite 1D systems. This algorithm
differs from the iTEBD algorithm\cite{cond-mat/0605597} in that
it uses a variational approach instead of imaginary time evolution,
and from the PWFRG\cite{Nishino:1995kx} in that it considers an
infinite system from the start. Furthermore, since the algorithm
itself employs matrix product operators, it has the important
advantage of being capable of modeling long range interactions, and
in particular any interaction which can be approximated by a sum of
decaying exponentials. In order to benchmark the algorithm, we have
computed an infinite MPS for the ground state of the Haldane-Shastry
model. The corresponding two-point correlators are in remarkable
agreement with the exact solution up to distances above a thousand
spins.

In conclusion, our results indicate that this algorithm adds
significantly to the existent tools to address 1D many-body systems
since it allows the properties of bulk-scale materials to be studied
for realistic long-range potentials. Furthermore, it admits a natural
extension to lattice systems in higher spatial dimensions, for which
work is currently in progress.

We note that upon completion of this work, we learned of simultaneous
work on an equivalent algorithm by Ian McColloch in
Ref. \onlinecite{0804.2509}; his presentation includes a detailed
comparison of the convergence of the
iTEBD\cite{cond-mat/0605597} and the variational approaches.

\acknowledgements{
Gregory Crosswhite performed this work as part of the East Asia Pacific Summer Institute program, cofunded by the National Science Foundation and the Australian Academy of Science;  additional support was received from the Computational Science Graduate Fellowship program, U.S. Department of Energy Grant No. DE-FG02-97ER25308.  Andrew Doherty and Guifr\'e Vidal (FF0668731) acknowledge support from the Australian Research Council.}

\bibliography{references}

\end{document}